\documentclass[reprint,floatfix,notitlepage,superscriptaddress]{revtex4-1}

\usepackage{hyperref}
\usepackage{graphicx}
\usepackage{xcolor}
\usepackage{multirow}

\usepackage[caption=false]{subfig}

\usepackage{bm}

\makeatletter
\@ifundefined{date}{}{\date{}}
\makeatother

\usepackage{amsmath, amsfonts, amssymb}
\usepackage{mathtools}
\usepackage{physics}
\usepackage{bbm}
\usepackage{bm}
\usepackage{pifont} 

\usepackage{soul} 

\newcommand{\ncmd}{\newcommand}
\ncmd{\nn}{\nonumber}
\ncmd{\half}{\frac{1}{2}}
\ncmd{\mbf}[1]{\bs{#1}}
\ncmd{\gam}{\gamma}
\ncmd{\sig}{\sigma}
\ncmd{\pha}{\alpha}
\ncmd{\lam}{\lambda}
\ncmd{\dl}{\delta}
\ncmd{\kap}{\kappa}
\ncmd{\eps}{\epsilon}
\ncmd{\Lam}{\Lambda}
\ncmd{\Gam}{\Gamma}
\ncmd{\Dl}{\Delta}
\ncmd{\Ups}{\Upsilon}
\ncmd{\Om}{\Omega}
\ncmd{\om}{\omega}
\ncmd{\veps}{\varepsilon}
\ncmd{\vphi}{\varphi}
\ncmd{\vtheta}{\vartheta}
\ncmd{\tw}{\text{w}}
\ncmd{\pll}{\parallel}
\ncmd{\mc}{\mathcal}
\ncmd{\mf}{\mathfrak}
\ncmd{\bs}{\boldsymbol} 
\ncmd{\trans}[1]{{#1}^\intercal}
\ncmd{\eq}[1]{Eq. \eqref{#1}}
\ncmd{\fig}[1]{Fig. \ref{#1}}
\renewcommand{\var}[1]{\text{var}(#1)}
\ncmd{\mrm}[1]{\mathrm{#1}}

\ncmd{\suppl}{\note{`Supplementary Information'}}
\ncmd{\bc}{\text{BC}}
\ncmd{\pd}[1]{\partial_{#1}}

\definecolor{blue2}{rgb}{0.2, 0.2, 0.6}
\definecolor{blue3}{rgb}{0.16, 0.32, 0.75}
\definecolor{darkred}{rgb}{0.8,0,0}
\definecolor{royalblue}{rgb}{0.0, 0.14, 0.4}
\definecolor{magenta}{cmyk}{0,.9,0,0.2}
\definecolor{amethyst}{rgb}{0.6, 0.4, 0.8}
\definecolor{cadmiumgreen}{rgb}{0.0, 0.42, 0.24}
\definecolor{deepcarmine}{rgb}{0.66, 0.13, 0.24}
\definecolor{forestgreen}{rgb}{0.13, 0.55, 0.13}

\ncmd{\sect}[1]{\emph{\color{blue3}{#1}}~---~}

\ncmd{\para}[1]{\paragraph*{{\color{black}{\bf #1:}}} }

\ncmd{\note}[1]{{\color{magenta}{[\ding{168} #1}]}}

\begin{document}

\title{
Dual-use quantum hardware for quantum resource generation and energy storage
}

\author{Vaibhav Sharma}
\thanks{These authors contributed equally to this work.}
\affiliation{Smalley-Curl Institute, Rice University, Houston, TX 77005}
\affiliation{Department of Physics and Astronomy, Rice University, Houston, TX 77005}
\author{Yiming Wang}
\affiliation{Smalley-Curl Institute, Rice University, Houston, TX 77005}
\affiliation{Department of Physics and Astronomy, Rice University, Houston, TX 77005}
\affiliation{Extreme Quantum Materials Alliance, Rice University, Houston, TX 77005}
\author{Shouvik Sur}
\thanks{These authors contributed equally to this work.}
\affiliation{Smalley-Curl Institute, Rice University, Houston, TX 77005}
\affiliation{Department of Physics and Astronomy, Rice University, Houston, TX 77005}
\affiliation{Extreme Quantum Materials Alliance, Rice University, Houston, TX 77005}

\date{\today}
\begin{abstract}
Efficient generation of quantum resources is a central objective of modern quantum technological platforms.
Independently, quantum batteries have emerged as nanoscale devices that utilize collective quantum effects to store energy with a charging advantage over classical strategies. 
Here, we show a direct connection between these two pursuits: protocols for fast generation of resourceful quantum states can simultaneously charge a quantum battery with a collective advantage, and conversely, a quantum battery protocol with a charging advantage rapidly produces resource-rich states.
Using this connection, we propose an integrated hardware protocol on superconducting circuits in which each experimental run can interchangeably accomplish either quantum battery charging, or quantum sensing through generation of metrologically useful states.  
Our results establish that quantum resources and stored energy are distinct yet simultaneously-producible quantities within the same dynamics. This opens the door to polymorphic quantum architectures that dynamically switch between sensing and energy-storage functions, thereby producing additional functionalities without extra hardware cost.
\end{abstract}

\maketitle

\twocolumngrid

\para{Introduction} 
Quantum technologies draw their power from uniquely non-classical features of quantum mechanics to achieve advantages in computation, communication, and sensing~\cite{Nielsen2000,Deutsch2020}.
Entanglement, coherence, and non-Gaussianity have all been formalized as resources within the framework of quantum resource theories~\cite{Chitambar2019}, and their efficient generation on physical hardware underpins the performance of every branch of quantum technology.
For instance, quantum-enhanced metrology, that informs quantum sensing protocols, utilizes entanglement as a resource to surpass the standard quantum limit, achieving Heisenberg-limited precision in parameter estimation~\cite{Giovannetti2006,Giovannetti2011,Pezze2018,Degen2017,metrologyreview}.

Concurrently, the energy required to operate quantum hardware at scale has 
become a pressing concern, and motivates a unified approach that connects quantum 
thermodynamics and information science in order to ensure energy-efficient, sustainable quantum technologies~\cite{Auffeves2022}. 
Quantum batteries---nanoscale devices that store and release energy by 
harnessing collective quantum effects---have emerged as a natural arena where resource-theoretic and energetic perspectives 
converge~\cite{Alicki2013,Binder2015,Campaioli2024RMP,unknownstate}. 
Theoretical work has demonstrated that many-body or collective effects can yield super-extensive scaling of charging power which implies that the time required to charge a many-body quantum battery can be reduced with increasing number of battery units~\cite{Campaioli2017,Ferraro2018,Andolina2019extractable,Rossini2020SYK, Andolina2025, le2018,bosspaper,qbreview}.
This inverse relationship between battery-size and charging time constitutes a charging advantage. 
Experimental realizations of quantum batteries have now spanned multiple platforms, from organic semiconductor microcavities to superconducting transmon qubits, and together demonstrate that a quantum advantage is possible in energy storage~\cite{quach2022superabsorption, joshi2022, hymas2025experimental, hu2026}.

\begin{figure}[!t]
\centering
\includegraphics[width=0.95\columnwidth]{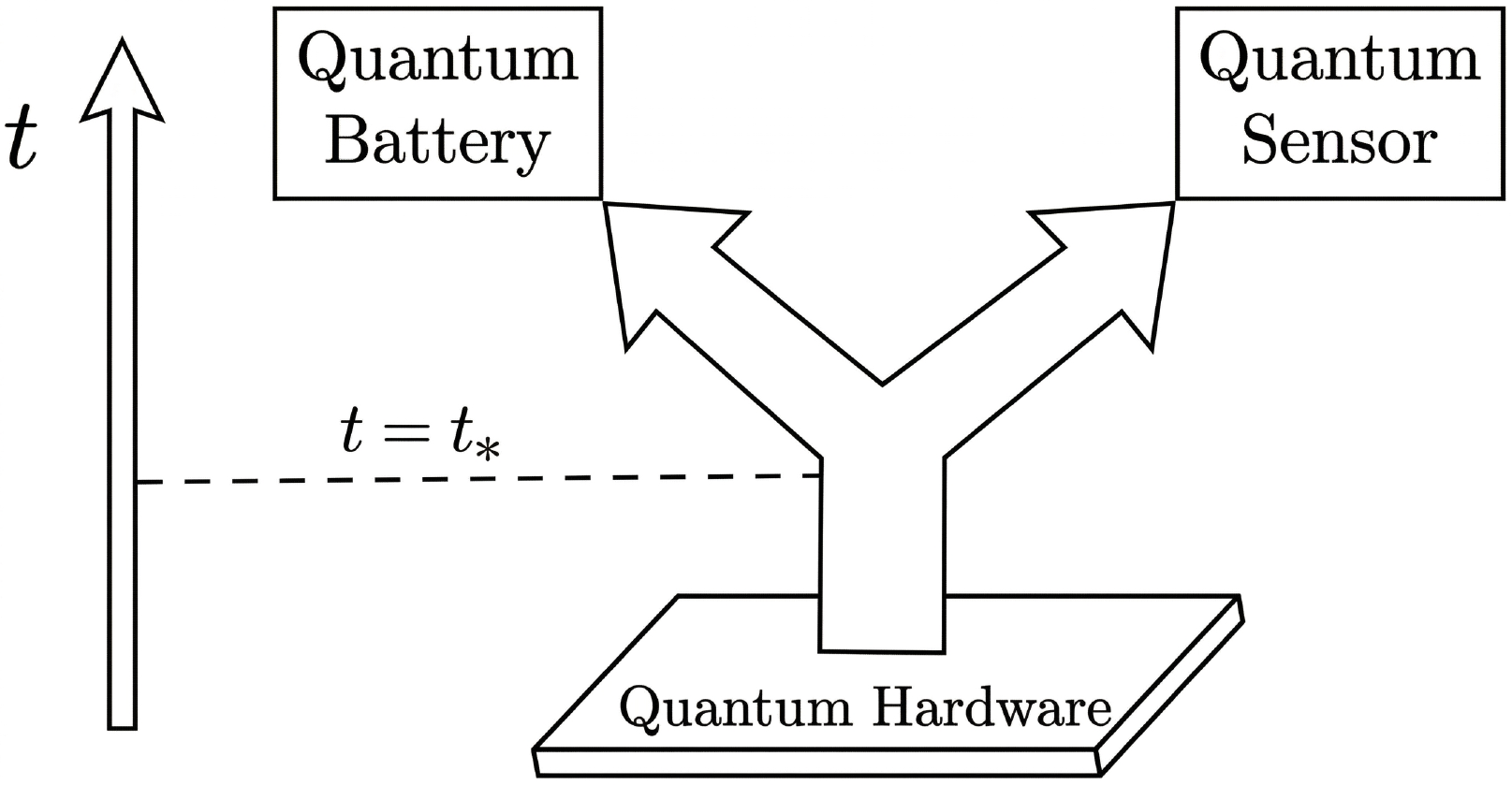}
\caption{The same quantum hardware can serve either as a quantum battery or a quantum sensor because the process of charging a quantum battery generates quantum resources. Here, we demonstrate a concrete protocol where the quantum hardware remains agnostic of its intended use until a time $t = t_*$ where a suitable metrological resource peaks. One can then choose to either time-evolve into a fully charged quantum battery or perform quantum sensing. 
As a bonus, after sensing, there exists a finite probability to obtain a fully charged battery.}
\label{fig:schematic}
\end{figure}

So far, achieving a charging advantage in quantum batteries and generation of resource-rich quantum states have developed as largely independent pursuits with connections between them remaining largely unexplored. 
However both fields share the common objective of speed: in quantum metrology, one  seeks to evolve an unentangled initial state into a metrologically useful target state in the shortest possible time through fast state-preparation protocols, thereby minimizing exposure to decoherence~\cite{Deffner2017, Guery-Odelin2019, Kitagawa1993,Ma2011}; quantum battery protocols seek to evolve a low-energy ground state to a high-energy excited state as rapidly as possible to achieve a charging advantage~\cite{Campaioli2024RMP}.
The speed-up in both cases stems from collective quantum effects, suggesting a deeper connection between them and  making their co-existence possible in a single physical set-up. Thus we can leverage this co-existence to develop multi-use protocols for quantum hardware where the same system can play multiple roles.

Towards this goal, we show that the same underlying Hamiltonian dynamics can accomplish the preparation of resourceful states and quantum battery charging, without requiring independently engineered protocols. 
We utilize this connection to propose \emph{dual-use quantum hardwares} that serve  interchangeably as a battery and a sensor, as presented schematically in Fig.~\ref{fig:schematic}. 
Within our protocol, rapid charging is facilitated by a dynamics that connects states in the Hilbert space that are vastly separated in energy, thereby generating strong energy-fluctuations and a concomitant enhancement in multipartite entanglement.
Utilizing this temporal structure, the protocol interrupts the charging trajectory to harvest quantum resources at their peak, before completing the charging cycle.

\para{State preparation protocols as charging quantum batteries} We begin by concretely demonstrating the connection between fast-state preparation protocols and efficient charging of  quantum batteries. 
In quantum state preparation, one starts with a simple initial state, $|\psi_i\rangle$, with the goal to evolve the system into a final state, $|\psi_f\rangle$, that exhibits strictly non-classical features such as topological order~\cite{topo1,topo2,topo3,topo4}, quantum computational utility~\cite{qc1,qc2} and enhanced metrological capability~\cite{squeezing1,squeezing2,squeezing3,squeezereview,spinsqueezingreview}. 
The preparation protocol can be written as~\cite{topo1,squeezing2,unitary1, nonunitary1,nonunitary2,nonunitary3,nonunitary4}
\begin{equation}
   |\psi_f \rangle = \hat V(t = t_f)|\psi_i\rangle,
\end{equation}
where $\hat V(t)$ is a time-dependent operator that transforms the initial state into the final state after evolution time $t_f$. 
Generally, the goal of any state preparation protocol is to minimize  $t_f$ in order to avoid the inevitable effects of decoherence and noise before $\ket{\psi_f}$ is utilized. 
This goal of fast state preparation connects it to the quest for charging advantage in quantum batteries.

To map this to a quantum battery, we interpret $\ket{\psi_i}$ as the uncharged initial state of a quantum battery. 
As $|\psi_i\rangle$ is typically chosen to be an easy-to-prepare state, such as a product state of spins, we can find a Hamiltonian, $\hat H_\mathrm{B}$, such that $|\psi_i\rangle$ is a low energy eigenstate of $\hat H_\mathrm{B}$. 
This makes $\hat H_\mathrm{B}$ our quantum battery Hamiltonian. 
The state preparation operator, $\hat V(t)$, can be simply mapped to the charging protocol of a quantum battery. 
Finally, $|\psi_f\rangle$ can be interpreted as the charged state of the battery as long as the following condition is satisfied:
\begin{equation}
   \Delta E = \langle \psi_f | \hat H_\mathrm{B} | \psi_f \rangle - \langle \psi_i | \hat H_\mathrm{B} | \psi_i \rangle > 0.
\end{equation}
This implies that the charging protocol $\hat V(t)$ has successfully  injected energy $\Delta E$ in time $t_f$. 
If  $t_f$ scales as $\Delta E^{-\eta}$ with $\eta >0$, then the charging power, $\Delta E / t_f$, exhibits a super-linear scaling which implies a quantum charging  advantage~\cite{quantumbatteryreview}.
Therefore, in the process of preparing the final state $|\psi_f\rangle$, we have produced both quantum resources and energy storage, and, depending on the state-preparation protocol, a charging advantage too.

{Although our mapping will allow several existing protocols for fast state-preparation to be interpreted as quantum batteries, a  charging advantage can be claimed only after additional constraints are imposed on the time-evolution operator $\hat V(t)$. 
We elucidate this for unitary protocols where $\hat V(t) = e^{-i \int_0^t dt' \hat H_c(t')}$ and $\hat H_c(t')$ is some charging Hamiltonian. 
The standard constraint is to ensure that the Mandelstam-Tamm quantum speed limit (QSL), controlling the evolution out of the initial state, remains invariant under changes in the charging protocol~\cite{bosspaper,Campaioli2017}. 
The QSL dictates that the minimum time needed for a state $|\psi_i\rangle$ to evolve into an orthogonal state is given by $t_{min} = \hbar \pi/(2\sqrt{\text{var}(\hat H_c)})$ where $\text{var}(\hat H_c) = \langle \psi_i |\hat H_c^2|\psi_i\rangle - (\langle \psi_i |\hat H_c|\psi_i\rangle)^2$. 
}

We exemplify the above with the help of 
 a state-preparation protocol for an ensemble of $N$ spin-$1/2$'s or qubits with $\hat H_c =  \chi  \qty(\sum_{n=1}^{N} \hat \sigma^x_n)^2 $, i.e. the one-axis twist (OAT) model, where $\hat \sigma^x_n$ is the Pauli operator acting on the $n^{\text{th}}$ spin. 
This model provides a well-known protocol for unitary preparation of states with high multipartite entanglement which are witnessed by both the quantum Fisher information (QFI) and spin-squeezing~\cite{witness1,witness2,witness3,Fadel2025}. Such states enable precise measurements that can surpass the standard quantum limit imposed by the Heisenberg uncertainty principle~\cite{squeezereview,spinsqueezingreview}. 
Initializing the system in a coherent spin state, the optimal time to prepare a squeezed state is given by $t_f' \sim N^{-2/3}/\chi$~\cite{Kitagawa1993,spinsqueezingreview}. 
Thus the OAT model is a fast preparation protocol since $t_f'$ decreases with increasing $N$.

To evaluate if this model also exhibits a charging advantage, we 
impose the QSL constraint in the ground state of the battery Hamiltonian, $H_\textrm{B} = \omega_0 \sum_n^N \hat \sigma_n^z$. 
Relegating the details to the Supplementary Materials (SM)~\cite{sm}, we only highlight the key results here.
We find that, while $\Delta E \propto N$, the charging time $t_f$ is proportional to $N^{-1/4}$ leading to a superextensive charging power proportional to $N^{5/4}$. 
This shows that the OAT model can be used for both fast preparation of a metrologically useful state and fast charging of a quantum battery. 
We note that the OAT model has previously been analyzed as the Lipkin-Meshkov-Glick (LMG) quantum battery model and it was shown to exhibit a charging advantage~\cite{lmg1,lmg2}. Our results here add the consequences of imposing the QSL constraint to this model.

\para{Metrologically useful states from a quantum battery}
Just as fast state preparation protocols can be re-interpreted as fast charging of a quantum battery, so too can  well-known quantum battery models with charging advantage provide templates for fast preparation of resourceful quantum states. 
In particular, the same collective quantum processes that produce charging-advantage in quantum batteries~\cite{bosspaper} also build multipartite entanglement, making them well-suited to produce useful quantum states while charging the battery.
Because of their experimental  relevance~\cite{quach2022superabsorption, joshi2022, hymas2025experimental, hu2026}, we will consider the initial uncharged state of the quantum battery, $|\psi_i\rangle$, to be unentangled and the ground state of the battery Hamiltonian. 
The charged state $|\psi_f\rangle$ is some high energy state within the spectrum of $\hat H_B$ that is orthogonal (or nearly orthogonal) to $|\psi_i\rangle$.
If the charging Hamiltonian $\hat H_\mathrm{AB}$ directly couples these two states, such that $\langle \psi_f|\hat H_\mathrm{AB}|\psi_i\rangle \neq 0$, 
then we get a quantum battery model with the charging time $t_f \sim \Delta E^{-\eta}$ with $\eta > 0$, leading to a charging advantage~\cite{bosspaper}.

In general, this charging process generates states at intermediate times that are of the form $\ket{\psi(t)} = \alpha(t) |\psi_i\rangle + \beta(t) |\psi_f\rangle + \ldots$ where $\alpha(t),\beta(t)$ are complex-valued  functions of $t$ and the ellipses represent additional terms that maybe present.
Owing to the superposition between $\ket{\psi_i}$ and $\ket{\psi_f}$, the width of the probability distribution of the eigenvalues of $\hat H_\mathrm{B}$ in {$\ket{\psi(t)}$} would be proportional to $\Delta E$. 
Provided the dynamics is dominated by superpositions between sectors of $\hat H_\mathrm{B}$ separated by $O(\Delta E)$ and $\Delta E$ is an extensive quantity, 
we observe that $|\psi(t)\rangle$ encodes super-extensive fluctuations of the operator $\hat H_\mathrm{B}$, leading to an $\Delta E^2$ scaling of $\var{\hat H_\mathrm{B}}$. 
Since the QFI for an operator $\hat O$ in a pure state $|\psi\rangle$ is given by $F_Q = 4  \var{\hat O}$~\cite{Fadel2025}, $F_Q(\hat H_\mathrm{B}) \propto \Delta E^2$, making  $\ket{\psi(t)}$ suitable for Heisenberg-limited sensing. 
Since this metrologically useful entanglement is not engineered independently, but emerges during the same dynamics that optimize rapid energy storage, this identification links  quantum advantage in batteries to quantum resource generation. 
We note that the entanglement generated in such dynamics is generally not a monotonic function of time or the instantaneous charging power~\cite{powerentanglementrelation}.

As a concrete and experimentally relevant demonstration of the above connection, we consider a quantum battery composed of a pair of coupled superconducting LC resonators with a fixed ratio of frequencies, as introduced in Ref.~\cite{Andolina2025}.
The quantum battery Hamiltonian for this system is well-approximated by (see the SM~\cite{sm} for details)
\begin{align} 
& \hat H(t) = \hat H_\mathrm{A} + \hat H_\mathrm{B} +  \lambda(t) \hat H_\mathrm{AB},   \nn \\
& \mbox{with} \quad  \hat H_\mathrm{A} = n\omega_0 \hat a^\dagger \hat a; \quad 
\hat H_\mathrm{B} = \omega_0 \hat b^\dagger \hat b; \nn \\
& \qquad \quad \hat H_\mathrm{AB} = g_n (\hat  a^\dagger \hat b^n + \mbox{h.c.}),
\label{eq:battery}
\end{align}
where $\hat H_\mathrm{B}$ ($\hat H_\mathrm{A}$) is the  battery (charger) Hamiltonian which corresponds to a bosonic mode having excitation frequency $\omega_0$ ($n \omega_0$), $n$ is a positive integer, and  $g_n$ controls the coupling between the two bosonic modes. 
The parameter $\lambda(t)$ is a time-dependent function such that $\lambda(t) = 1$ when $0<t<t_f$ and zero otherwise. 
During the process of charging in the interval $0<t<t_f$, the charging Hamiltonian $\hat H_\mathrm{AB}$ transfers energy from the charger mode to the battery mode by converting one quantum from mode $A$ to $n$ quanta in mode $B$. 
The stored energy in the battery is given by $E_B = \omega_0\langle \hat b^\dagger \hat b \rangle$. 

Due to the invariance of $\hat H$ under $\hat a \to e^{i n \theta} \hat a$ and $\hat b \to e^{i  \theta} \hat b$, $\hat Q = n \hat a^\dagger \hat a + \hat b^\dagger \hat b$ is conserved during charging. 
Therefore, if the charger-battery ensemble is initiated in a state with a fixed $Q \equiv \expval{\hat Q}$, it cannot leave this sector during charging.
Further, the resonance condition that fixes the ratio between the frequencies of the resonators also implies  $[\hat H_\mathrm{AB}, \hat H_\mathrm{A} + \hat H_\mathrm{B}] = 0$, which preserves the total energy of the charger-battery ensemble as the battery is charged. 

In order to exhibit the joint production of charging and metrological resources, we  consider initial states of the form $\ket{\psi_i} = \ket{\psi(t=0)} = \ket{1}_A \ket{N_B}_B$ with $0 \leq N_B < n$, which results in $Q = n + N_B$. 
The non-triviality of the charging process results from the $\hat H_\mathrm{AB}$-mediated direct coupling between $\ket{\psi_i}$  and the final or fully-charged state $|\psi_f\rangle = \ket{0}_A \ket{N_B + n}_B$, which results in $\Delta E = n \omega_0$ and the matrix element  $\langle \psi_f|\hat H_\mathrm{AB}|\psi_i\rangle = \Omega_Q$ with $\Omega_Q =\sqrt{n!} \qty(\prescript{Q\mkern-0.5mu}{}C_{n})^{1/2}    g_n$. 
Thus, in dynamics, we get an effective two-level system with  $\hat H_\mathrm{AB}$ causing Rabi oscillations between $|\psi_i\rangle$ and $|\psi_f\rangle$, such that the state in the charging interval is given by 
\begin{align}
|\psi(t)\rangle = 
e^{-iE_Q t}\left[\cos(\Omega_Q t )\ket{\psi_i} - i \sin (\Omega_Q t) |\psi_f\rangle \right].
\end{align}
where $E_Q = \omega_0 Q$.
At the end of the charging cycle, we have $t = t_f \equiv \pi/(2\Omega_Q)$ and $|\psi(t_f)\rangle = |\psi_f\rangle$ upto a global phase.
Establishing a charging advantage requires a fair comparison between time evolution under the non-linear Hamiltonian $\hat H(t)$ and its linear counterpart ($n=1$) using the Mandelstam-Tamm QSL constraint~\cite{Campaioli2017, Andolina2025}.
Here, this constraint relates $g_n$ and  $g_1$,  and allows the Rabi frequency to be expressed in terms of $g_1$ such that $\Omega_Q = g_1  \sqrt{n + (2n + 1)N_B} $. 
Since the rate of charging is controlled by  $\Omega_Q^{-1}$, $t_f \sim n^{-1/2}$ at a fixed $N_B$, implying an $\eta = 1/2$ and, thus, a charging advantage.
We note that in the limit $N_B = 0$ we reproduce the results in Ref.~\cite{Andolina2025}.

\begin{figure}
\centering
\includegraphics[width=0.95\columnwidth]{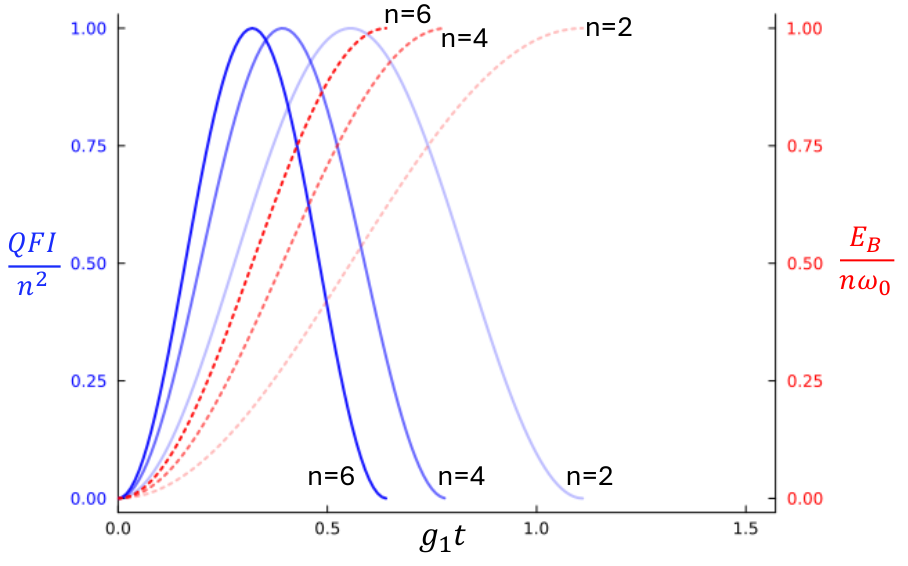}
\caption{Stored battery energy $E_B/(n\omega_0)$ (red dotted lines) and QFI normalized by $n^2$ (blue solid lines) during the charging dynamics as a function of time $t$ for different values of $n$ (larger $n$ corresponds to darker shades) in the quantum battery model in Eq.~\ref{eq:battery} with $Q = n$. Increasing $n$ leads to a decrease in the battery charging time. The QFI peaks earlier than the time to fully charge the battery and the peak time also reduces with increasing $n$.
}
\label{fig:chargingdynamics}
\end{figure}

At intermediate times, $0 < t < t_f$, the charging process generates QFI with respect to the operator $\hat n_\mathrm{B} = \hat b^\dagger \hat b$, with the QFI peaking at $t = t_1 \equiv t_f/2$.
Explicitly, at $t  = t_1$, the quantum state is
\begin{equation}
\ket{\psi_1} \equiv \ket{\psi(t_1)} = \frac{e^{-i E_Q t_1}}{\sqrt{2}} \qty(\ket{1}_A \ket{N_B}_B - i \ket{0}_A \ket{N_B + n}_B ),
\end{equation}
and the QFI  is  $F_Q(\hat n_B) = n^2$. 
Since $F_Q(\hat n_B)$ scales quadratically with $n$, $\ket{\psi_1}$ is capable of Heisenberg-limited sensing. 
We note that $\ket{\psi_1}$ is {analogous to} the well-known NOON state that can provide precise phase measurements~\cite{noon}. 
The generation of high QFI here comes from the collective charging process of the quantum battery which is producing useful quantum resources along the way. 
{Since the preparation time $t_1$ scales as $1/\Omega_Q$}, we also get a  fast preparation of $\ket{\psi_1}$. {Fig.~\ref{fig:chargingdynamics} shows the stored energy and QFI during the charging dynamics for a few different values of $n$. 
We can see that the time to fully charge the battery reduces with increasing $n$~\cite{Andolina2019extractable}. 
The QFI peaks in the middle of this charging process and this peak consequently shifts to smaller times as $n$ increases.}
In closing, we note that this charging dynamics can also generate two-mode squeezed states.
In order to access this distinct metrological resource, however, $\ket{\psi_i}$ needs to be a coherent state, as detailed in the SM~\cite{sm}.

\para{Hardware setup combining metrology and charging}
\begin{figure}
\centering
\includegraphics[width=0.95\columnwidth]{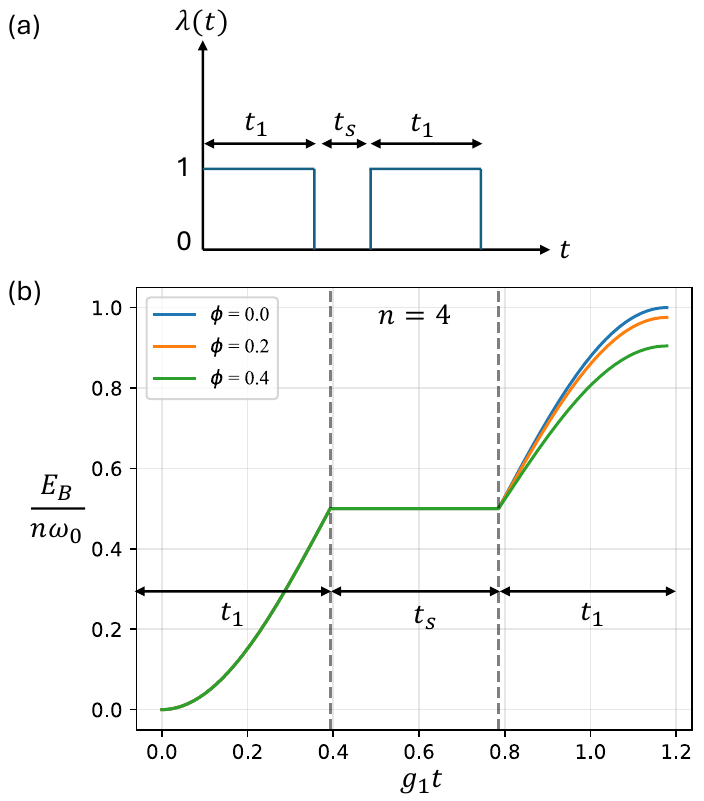}
\caption{(a) Protocol combining charging (for time $2t_1$ with $\lambda(t)=1$) and sensing (for time $t_s$ with $\lambda(t) = 0$) using the two coupled superconducting LC resonator based quantum battery model in Eq.~\ref{eq:battery}. (b) Stored energy in the battery $E_B/(n\omega_0)$ during the protocol in (a) for $n=4$ with three different values of the sensing parameter $\phi$ ($\phi = 0.0,0.2,0.4$) in increasing order from top to bottom. The final stored energy decreases as $\phi$ increases. 
}
\label{fig:sensingprotocol}
\end{figure}

We have now seen how charging advantage in quantum batteries and resourceful quantum state preparation protocols can coexist on the same quantum hardware. 
Inspired by this, we propose a proof-of-concept hardware setup based on the coupled LC resonators where these two functionalities can be combined.

For the present purpose we promote the temporal protocol $\lambda(t)$ in Eq.~\eqref{eq:battery} to a more non-trivial form, as shown in Fig.~\ref{fig:sensingprotocol}(a). 
Following it, we start with an initial uncharged state, $|\psi_i\rangle = |1\rangle_A |0\rangle_B$ at time $t=0$. 
We first  evolve the state under the charging Hamiltonian, $\hat H_\mathrm{AB}$, until time $t_1$ such that we get the state $|\psi_1\rangle$ possessing high QFI. 
We then switch off $\hat H_\mathrm{AB}$ and perform quantum sensing. 
Since  now $F_Q(\hat n_B) = n^2$, 
 $|\psi_1\rangle$ is a suitable state to sense a parameter $\phi$ encoded in a Hamiltonian given by $\hat H_s = -\phi \hat n_b$. 
 The parameter $\phi$ corresponds to an external flux through the superconducting resonator~\cite{fluxtunablelc}. It gets imprinted as a phase onto the post-sensing state $|\psi_s\rangle$. We then resume the charging dynamics under $\hat H_{AB}$ for time $t_1$ to get the final state $|\psi_m\rangle$. The imprinted phase can be read off by projectively measuring
 $\hat n_a = \hat a^\dagger \hat a$. This measurement leaves the battery partially charged with the stored energy dependent on $\phi$ as shown in Fig.~\ref{fig:sensingprotocol}(b). 

We will now show how the final stored energy quantitatively depends on the sensing parameter $\phi$. 
The post-sensing state $|\psi_s\rangle$ with the imprinted phase $\phi$ is given by
\begin{equation}
\ket{\psi_s} \equiv e^{-i \hat H_s t_s}\ket{\psi_1}  
\propto (\ket{1}_A \ket{0}_B +  e^{i\phi nt_s} \ket{0}_A \ket{n}_B). 
\end{equation}
The final state $|\psi_m\rangle$ obtained after resuming the charging dynamics for time $t_1$ is
\begin{align}
\ket{\psi_m}  \propto - \sin{\frac{\phi n t_s}{2}} \ket{1}_A \ket{0}_B +  \cos{\frac{\phi n t_s}{2}} \ket{0}_A \ket{n}_B.
\end{align}
The projective measurement of $\hat n_a$ in state $|\psi_m\rangle$ gives $\expval{\hat n_a} = 1$ with probability, $p_1 =\sin^2 (\phi n t_s/2)$, and $\expval{\hat n_a} = 0$ with probability, $p_0 = \cos^2 (\phi n t_s/2)$. 
Measuring these probabilities through multiple runs of the experiment helps us estimate the parameter $\phi$. 
Note that the probabilities are enhanced by a factor of $n$ due to the high QFI present in the sensing state $|\psi_1\rangle$. {This procedure is the familiar Ramsey-interferometry process one would use to read out the phase $\phi$ from a NOON state. In our case, there is an additional interpretation of battery charging attached to it.}

The post-measurement ensemble of the quantum battery is described by the density matrix, $\rho$ where
\begin{equation}
    \rho = \sin^2 \frac{n\phi t_s}{2}|0_B\rangle \langle0_B| + \cos^2 \frac{n\phi t_s}{2}|n_B\rangle \langle n_B|.
\end{equation}
 The average stored energy of this ensemble is given by $n\omega_0 \cos^2 |n\phi t_s/2|$. When $|n \phi t_s/2| < \pi/2$, the stored energy monotonically decreases as a function of $\phi$. This can be seen in Fig.~\ref{fig:sensingprotocol}(b).
If we are sensing a small parameter such that $|\phi| n t_s/2 \ll 1$, we get a nearly fully-charged quantum battery ensemble even after using the system for quantum sensing. 
The ergotropy, which determines the maximum extractable work from the battery~\cite{quantumbatteryreview}, 
then becomes $n \omega_0 \cos n \phi t_s$, comparable to the case where we do not interrupt the charging cycle for sensing.

\para{Conclusion and Outlook} 
Our work lays the foundation for building multi-use quantum technological platforms by adding an independent direction in the pursuit of on-chip energy source for quantum hardware~\cite{Kurman2026}.
In particular, we have  shown how the charging advantage in a quantum battery model sets the amount of metrologically useful entanglement that can be generated from the 
same underlying dynamics, linking the speed of energy storage to quantum sensing precision.
Using a model of two coupled superconducting LC resonators with commensurate mode frequencies, we showed that the non-linear charger-battery interaction generates states at intermediate times with super-extensive scaling of QFI  and two-mode squeezing, depending on the choice of the initial state. We proposed a hardware protocol in which each experimental run on this device can interchangeably  accomplish either quantum sensing or quantum battery charging. 

Broadly, the notion of dual-use developed here is not restricted to LC resonator architectures: any platform supporting non-linear  bosonic interactions or collective spin dynamics, eg.  trapped ions~\cite{nonlinearsqueezing}, photonic systems~\cite{photonicskerr}, and solid-state spin ensembles~\cite{spinensemble}, could host analogous protocols, enabling a comparative benchmarking.
Our results on dual-use quantum hardware for sensing and energy storage, combined with recent related works~\cite{allen2025quantum, Kurman2026}, enables an extension into a multi-use quantum hardware paradigm, where quantum sensing, quantum computing, and energy storage maybe be achieved on the same hardware platform.

\acknowledgements
We thank Vir Bulchandani, Diego Fallas Padilla, and Andrea Rocchetto for helpful feedback on our manuscript.
V.S. acknowledges support from the J. Evans Attwell Welch fellowship by the Rice Smalley-Curl Institute.
S.S. and Y.W. have been supported in part by the NSF Grant No. DMR-2220603.


\bibliography{references.bib}

\appendix

\renewcommand{\theequation}{S\arabic{equation}}
\renewcommand{\thefigure}{S\arabic{figure}}

\onecolumngrid

\section{Derivation of the superconducting LC resonator quantum battery model}\label{app:transmons}

Following the arguments in Ref.~\cite{Andolina2025}, we reproduce here the derivation of the quantum battery model in the main text given by
\begin{equation}\label{speq:battery}
    \hat H_{\text{model}} = n\omega_0 \hat a^\dagger \hat a + \omega_0 \hat b^\dagger \hat b + g_n (\hat a^\dagger \hat b^n + \text{h.c.})
\end{equation}

The model above can be obtained by considering two-coupled superconducting LC resonators with inductance $L_i$ and capacitance $C_i$ where $i \in \{1,2\}$. These resonators can be expressed as bosonic modes annihilated by mode operators $a,b$ respectively. The mode frequencies are $\omega_i = 1/\sqrt{L_i C_i}$. We consider them to be coupled by a Josephson junction. The Hamiltonian that models this system is given by
\begin{equation}\label{eq:twotransmon}
    \hat H = \omega_1 \hat a^\dagger \hat a + \omega_2 \hat b^\dagger \hat b + E_J \cos (\hat\phi_1 - \hat\phi_2).
\end{equation}

Here $\hat\phi_1 = \lambda_1 (\hat a + \hat a^\dagger)$ and $\hat \phi_2 = \lambda_2 (\hat b + \hat b^\dagger)$ are bosonic mode quadratures, corresponding to the quantized flux degree of freedom. The coupling $E_J$ can be tuned by an external flux in the Josephson junction and $\lambda_i = 2e^2 \sqrt{L_i/C_i}$ where $e$ is the elementary electric charge. When $\lambda_i \ll 1$, we can expand the cosine term in Eq.~\ref{eq:twotransmon} as
\begin{equation}\label{eq:cosine}
    \cos (\hat\phi_1 - \hat\phi_2) = \sum_{m=0}^{\infty} \sum_{k=0}^{2m} \frac{(-1)^{3m-k}}{k! (2m-k)!} \hat\phi_1^k \hat\phi_2^{2m-k}.
\end{equation}

We can choose $\omega_1 = n\omega_0$ and $\omega_2 = \omega_0$. The coupling term can create and destroy excitations in the two bosonic modes. When $\lambda_i E_J \ll \omega_i$, only resonant terms in the expansion of the cosine in Eq.~\ref{eq:cosine} are important for dynamics. These resonant terms are those that commute with $\omega_1 a^\dagger a + \omega_2 b^\dagger b$. They correspond to creation of one excitation in the $a$ mode while destroying $n$ excitations in the $b$ mode and vice-versa. The leading order term satisfying these conditions comes from the term proportional to $\hat \phi_1 \hat\phi_2^{n}$ from the expansion in Eq.~\ref{eq:cosine}, corresponding to $k=1$ and $m = (n+1)/2$. Retaining only this leading order term, the model Hamiltonian becomes
\begin{equation}
    \hat H = n\omega_0 \hat a^\dagger \hat a + \omega_0 \hat b^\dagger \hat b + E_J \lambda_1 \lambda_2^n\frac{(-1)^{(3n+1)/2}}{n!} (\hat a(\hat b^\dagger)^n + \hat a^\dagger (\hat b)^n).
\end{equation}

This completes the derivation of the battery model Hamiltonian in Eq.~\ref{speq:battery} where $|g_n| = E_J \lambda_1 \lambda_2^n/n!$. 

\section{Charging advantage in one-axis twisting squeezing model using the Mandelstam-Tamm quantum speed limit}

Here we show that the one-axis twisting (OAT) squeezing Hamiltonian can be interpreted as a quantum battery with a charging advantage after a proper imposition of the Mandelstam-Tamm quantum speed limit (QSL) constraint. The OAT model is defined as
\begin{equation}
\label{eq:oneaxis}
\hat H_{oa} =  \omega_0 \hat S_z + \chi (\hat S_x^2),
\end{equation}
where we have considered a system of $N$ spin-1/2 particles and the collective spin operators are defined as $\hat S_\alpha = \sum_{n=1}^{N} \hat \sigma^\alpha_n$. 
The operators $\hat \sigma^\alpha_n$ are Pauli operators corresponding to the $n^{\text{th}}$ spin and $\alpha \in {x,y,z}$. For both the preparation of squeezed states as well as quantum battery charging, we consider the initial state to be a coherent spin state that has all spins aligned along the $-z$ axis, represented as $|\psi_0\rangle = \prod_n \otimes |\downarrow_n\rangle $. This setting can be interpreted as a quantum battery of $N$ spin-1/2 particles with a battery Hamiltonian given by
\begin{equation}
    \hat H_B = \omega_0 \hat S_z
\end{equation}
 and the charging Hamiltonian given by
 \begin{equation}
     \hat H_c = \chi \hat S_x^2.
 \end{equation}

The initial state $|\psi_0\rangle$ is the fully uncharged state of the battery with energy $E_0 = -N\omega_0/2$. This quantum battery model has been studied in the literature as the Lipkin-Meshkov-Glick (LMG) quantum battery model~\cite{lmg1,lmg2}. In these works, it was found that the stored energy $\Delta E$ is proportional to $N$ while the time to charge $t_f$ is proportional to $N^{-1/2}/\sqrt{\chi \omega_0}$~\cite{lmg2} leading to a superextensively scaling charging power given by $\Delta E/t_f \propto N^{3/2}$. This superextensivity with the number of battery units $N$ is what gives this quantum battery model a charging advantage. Here we show that one must impose the Mandelstam-Tamm QSL constraint on $\hat H_c$ to properly determine if there is a charging advantage as one increases $N$.

The QSL constraint by Mandelstam-Tamm states that the time to evolve between two given states $|\psi_i\rangle$ and $|\psi_f\rangle$ under a Hamiltonian $\hat H_e$ is lower bounded by
\begin{equation}\label{speq:qsl}
    t^e_{\text{min}} = \frac{\hbar \cos^{-1}(\langle \psi_f|\psi_i\rangle)}{\sqrt{\text{var}(H_e)}}.
\end{equation}

Here $\text{var}(H_e) = \langle \psi_i|\hat H_e^2|\psi_i\rangle - (\langle \psi_i|\hat H_e|\psi_i\rangle)^2$. To fairly compare two different charging protocols, one must ensure that the QSL mandated $t^e_{\text{min}}$ must be the same for both. A charging advantage corresponds to the case when a protocol's actual charging time can get closer and closer to this time $t^e_{\text{min}}$. This ensures that there is no speed up simply due to a larger coupling constant or an atypical initial state. In our case, we need to determine if the OAT model has a charging advantage when we increase the number of spins that are collectively coupled. Thus we impose the constraint that $t^e_{\text{min}}$ is equal for all such cases.  

Using Eq.~\ref{speq:qsl} with $|\psi_i\rangle = |\psi_0\rangle$, $\hat H_e = \hat H_c$ and $|\psi_f\rangle$ chosen as some charged state orthogonal to $|\psi_0\rangle$, we get in the limit of $N \gg 1$ 

\begin{equation}
    t_{\text{min}} = \frac{\sqrt{2}\pi\hbar}{\chi N}.
\end{equation}

We contrast this with the case where the $N$ spins are divided into $N/2$ disjoint sets containing two spins each where only the pair of spins within each set are collectively coupled. The charging Hamiltonian $\hat H_c '$ becomes
\begin{equation}
    \hat H_c' = \chi'\sum_{m=1}^{N/2} (\sigma^x_{2m-1}+\sigma^x_{2m})^2.
\end{equation}

The battery Hamiltonian $\hat H_B$ and the initial uncharged state remain the same, but the numer of collectively coupled spins have dropped from $N$ to $2$. Using Eq.~\ref{speq:qsl} to find the QSL under $\hat H_c'$, we get
\begin{equation}
    t'_{\text{min}} = \frac{\pi\hbar}{\chi' \sqrt{N}}.
\end{equation}

So far, there is no relation between $\chi'$ and $\chi$. We can get it by imposing that the two scenarios have the same QSL, that is, $t_{\text{min}} = t'_{\text{min}}$. Thus we get $\chi = \chi'/\sqrt{N}$. This shows that for OAT models, charging the number of collectively coupled spins should be compensated by a rescaling of the coupling $\chi$ by a factor of $\sqrt{N}$ so that a charging advantage can be established as a function of $N$.

We can now revisit the LMG quantum battery results studied in~\cite{lmg1,lmg2}. With the new rescaling of $\chi \to \chi/\sqrt{N}$, the charging time now becomes $t_f \sim N^{-1/4}/\sqrt{\chi \omega_0}$. The charging power is thus proportional to $N^{5/4}$, retaining a superextensive scaling and hence a charging advantage. Thus we have shown that the OAT model can be interpreted as a quantum battery, and a charging advantage can be established after imposing the QSL constraint.

\section{Mapping the bosonic model to a spin system} \label{app:collective-spin}
Here, we discuss a mapping between the bosonic system studied in the main text and a system to two flavors of spins that are collectively coupled.
We consider two ensembles of spin-$1/2$'s, $\mc A$ and $\mc B$.
The spins in the two groups interact through a collective and generalized $XY$ type interaction $~\sim [S_A^{+} (S_B^-)^n + \mbox{h.c.}]$---for $n=1$ the interaction is a combination of conventional $XY$ and a spin-orbit-like term.
We note that this is a generalized version of $n$-axis twisting Hamiltonian, where the twisting term is linearly coupled to another collective spin.
We include transverse magnetic fields to make this model fully quantum,
\begin{align}
H_{\mathrm{spin}} = - h_A S_A^z - h_B S_B^z + J_n [S_A^{+} (S_B^-)^n + \mbox{h.c.}].
\end{align}

In order to connect with the non-linear bosonic model in the main text, we ``bosonize'' this spin model by introducing the Holstein-Primakoff transformation, $S_A^z = N_A/2 - a^\dagger a; \quad S_A^+ = a \sqrt{N_A - a^\dagger a }$ (and the same for the spins in ensemble $\mc B$) with $N_A$ ($N_B$) being the number of spins in $\mc A$ ($\mc B$).
Assuming $N_A, N_B \gg 1$, upto a overall shift in energy, we can thus express the Hamiltonian as
\begin{align}
H_{\mathrm{spin}} \simeq h_A a^\dagger a  + h_B b^\dagger b + \sqrt{N_A N_B^n} J_n \qty[a^\dagger b^n + \mbox{h.c.} ].
\end{align}
It is now straightforward to identify the couplings in $H_{\mathrm{spin}}$ with those appearing in the non-linear bosonic model in the main section. 
We note that while $n$ is a shared parameter between the two models, $N_A$ and $N_B$ are parameters that are physically meaningful only in the spin-model and, as we will see below, play a key role in determining inter-state transitions.  

A direct analysis of the spin model reproduces the salient features discussed in the main text. 
Here, for simplicity, we will set $N_A = N_B = N$ with $n \leq N$ and assume the resonance condition $h_A = n h_B$ to arrive at the following conclusions:
\begin{enumerate}
\item When the system is initialized in the product state $\ket{\psi_0} = \ket{N/2-1}_A \otimes \ket{N/2}_B$  (this is the first excited state at $J_n =0$) the Hamiltonian dynamics exhibits Rabi oscillation because $S_B^+ \ket{N/2}_B = 0 = S_A^+ \ket{N/2}_A$.
\item Because we are interested in identifying which features of a charging Hamiltonian lead to a charging advantage, we need to compare two charging protocols that offer the same initial speed for traversing the Hilbert space. 
This is ensured by equating the variance of the total Hamiltonian in the initial state within  the framework of quantum speed limit ($\tau \sim 1/\sqrt{\mbox{var}(H)}$).
Specifically, we equate  the variance for the model with a general $n > 1$ (initial state $\ket{\psi_0} = \ket{N/2-1}_A \otimes \ket{N/2}_B$) with that for $n=1$ (initial state $\ket{\psi_0} = \ket{N/2-n}_A \otimes \ket{N/2}_B$), which leads to the constraint, 
\begin{align}
J_n = \frac{J_1}{n! \sqrt{\prescript{N\mkern-0.5mu}{}C_{n-1}}}.
\end{align}
We note that this is different than Kac-normalization which guarantees extensivity of the total energy.
However, here, the extensivity does not help establish a fair comparison between distinct protocols for the task at hand. 
\item With the results above it is straightforward to obtain the frequency of the Rabi oscillation between $\ket{\psi_0}$ and $\ket{\psi_1} = \ket{N/2}_A \otimes \ket{N/2 - n}_B$ for the non-linear spin model,
\begin{align}
\Omega_n =  \sqrt{n} \times 2J_1 \sqrt{N(N-n+1)}.
\end{align}
The direct dependence of $\Omega_n$ on $\sqrt{n}$ underscores the fact that the non-linear interaction term connects the initial and final states of the quantum battery in 1 step---this is contrasted with the linear model ($n=1$) where the connection involves $n$ steps.
Since all time-scales of interest in this Hamiltonian  dynamics are obtained by setting $\Omega_n t \sim 1$,  such time scales necessarily scale inversely with  $\sqrt{n}$, thereby offering an improved performance with increasing $n$.
\end{enumerate}

As noted earlier, the spin-model also possess new physics that is absent in its pure bosonic analog. 
In particular, from the perspective of a model for a spin- or qubit-based quantum battery, a charging advantage in terms of $N$ is still present at $n=1$, as $\Omega_1 \propto N$.

\section{QFI generation at short times} \label{app:shorttimeqfi}
In the main section, in order to demonstrate the scaling of the quantum Fisher information (QFI) with the model parameter, $n$ we considered the initial state $\ket{1}_B \ket{N}_B$.
This choice has the virtue of allowing a non-perturbative solution of the dynamics at any $n$, as explained in the main text.
Here, we generalize the initial occupation of the A-mode to $N_A$ and demonstrate the presence of the same $n^2$ enhancement of the QFI, at least at short time scales. 
Thus, we consider the system is initially prepared in a Fock state: $\ket{\psi(0)} = \ket{N_A}_A \ket{N_B}_B \equiv \ket{N_A, N_B}$, and it evolves under  $\hat{H}$ within the interval $0 \leq t \leq t_0$,
\begin{align}
    \hat{H} &= \hat{H}_A + \hat{H}_B + \hat{H}_{AB} \\
    \hat{H}_A &= n\omega_0 \hat{a}^\dagger \hat{a} \\
    \hat{H}_B &= \omega_0 \hat{b}^\dagger \hat{b} \\
    \hat{H}_{AB} &= g_n (\hat{a}^\dagger \hat{b}^n + \hat{a} (\hat{b}^\dagger)^n)
\end{align}
The goal is to analytically determine the variance of the B-boson number operator, $\hat{n}_B  = \hat{b}^\dagger \hat{b}$, up to second order in time $\mathcal{O}(t^2)$.

We separate the Hamiltonian into a free part $\hat{H}_0 = \hat{H}_A + \hat{H}_B$ and an interaction part $\hat{H}_{AB}$. 
Because $[\hat{H}_0, \hat{H}_{AB}] = 0$, the exact time evolution operator factors as:
\begin{equation}
    \hat{U}(t) = \exp(-it\hat{H}) = \exp(-it\hat{H}_0) \exp(-it\hat{H}_{AB})
\end{equation}
Applying this to the initial state $\ket{N_A, N_B}$, which is an exact eigenstate of $\hat{H}_0$ with eigenvalue $E_Q = \omega_0(nN_A + N_B)$, gives:
\begin{equation}
    \ket{\psi(t)} = e^{-itE_Q} \exp(-it\hat{H}_{AB}) \ket{N_A, N_B}
\end{equation}
To evaluate the state up to $\mathcal{O}(t^2)$, we expand the interaction exponential:
\begin{equation}
    \exp(-it\hat{H}_{AB}) \approx \hat{I} - it\hat{H}_{AB} - \frac{t^2}{2}\hat{H}_{AB}^2
\end{equation}

Applying $\hat{H}_{AB}$ to the initial state yields transitions to two adjacent states. Let us define the transition amplitudes:
\begin{align}
    A_+ &= g_n \bra{N_A + 1, N_B - n} \hat{a}^\dagger \hat{b}^n \ket{N_A, N_B} = g_n \sqrt{N_A + 1} \sqrt{\frac{N_B!}{(N_B - n)!}} \\
    A_- &= g_n \bra{N_A - 1, N_B + n} \hat{a} (\hat{b}^\dagger)^n \ket{N_A, N_B} = g_n \sqrt{N_A} \sqrt{\frac{(N_B + n)!}{N_B!}}
\end{align}
to obtain 
\begin{equation}
    -it\hat{H}_{AB}\ket{N_A, N_B} = -it A_+ \ket{N_A + 1, N_B - n} - it A_- \ket{N_A - 1, N_B + n}
\end{equation}
Applying $\hat{H}_{AB}$ a second time introduces terms that return to the original state $\ket{N_A, N_B}$, as well as terms that transition further away ($\ket{N_A \pm 2, N_B \mp 2n}$). 
The full state up to $\mathcal{O}(t^2)$ is:
\begin{align}
    \ket{\psi(t)} \approx e^{-itE_Q} \Bigg[ & \left(1 - \frac{t^2}{2}(A_+^2 + A_-^2)\right) \ket{N_A, N_B} \nonumber \\
    & - it A_+ \ket{N_A + 1, N_B - n} - it A_- \ket{N_A - 1, N_B + n} \nonumber \\
    & + \mathcal{O}(t^2) \text{ terms orthogonal to both} \ket{N_A, N_B} \text{ and } \order{t} \text{ terms} \Bigg]
\end{align}

We check that the corresponding probabilities for occupying each state are:
\begin{align}
    P_0 &= \left| 1 - \frac{t^2}{2}(A_+^2 + A_-^2) \right|^2 \approx 1 - t^2(A_+^2 + A_-^2) \quad (\text{for } \ket{N_A, N_B}) \\
    P_1 &= |-it A_+|^2 = t^2 A_+^2 \quad (\text{for } \ket{N_A + 1, N_B - n}) \\
    P_2 &= |-it A_-|^2 = t^2 A_-^2 \quad (\text{for } \ket{N_A - 1, N_B + n})
\end{align}
which implies $P_0 + P_1 + P_2 = 1 + \mathcal{O}(t^4)$, ensuring unitarity at this order.

\subsection*{Variance of $\hat n_B$}
The expectation value ${\hat{n}_B}$ in the state $\ket{\psi(t)}$ is the weighted sum of eigenvalues:
\begin{align}
    \expval{\hat{n}_B}(t) &= P_0 N_B + P_1 (N_B - n) + P_2 (N_B + n) \nonumber \\
    &= [1 - t^2(A_+^2 + A_-^2)] N_B + t^2 A_+^2 (N_B - n) + t^2 A_-^2 (N_B + n) \nonumber \\
    &= N_B + n t^2 (A_-^2 - A_+^2)
\end{align}
Similarly, 
\begin{align}
    \expval{\hat{n}_B^2}(t) &= P_0  N_B^2 + P_1 (N_B - n)^2 + P_2 (N_B + n)^2 \nonumber \\
    &= [1 - t^2(A_+^2 + A_-^2)] N_B^2 + t^2 A_+^2 (N_B - n)^2 + t^2 A_-^2 (N_B + n)^2 \nonumber \\
    &= N_B^2 + 2n N_B t^2 (A_-^2 - A_+^2) + n^2 t^2 (A_+^2 + A_-^2) 
\end{align}
Therefore, the variance 
\begin{align}
    \Delta{\hat{n}_B}(t) &= \left[ N_B^2 + 2n N_B t^2 (A_-^2 - A_+^2) + n^2 t^2 (A_+^2 + A_-^2) \right]  - \left[ N_B^2 + 2 N_B n t^2 (A_-^2 - A_+^2) \right] \nonumber \\
    &= n^2 t^2 (A_+^2 + A_-^2) \nn \\
    &= n^2 t^2 g_n^2 \left[ (N_A + 1) \frac{N_B!}{(N_B - n)!} + N_A \frac{(N_B + n)!}{N_B!} \right].
\end{align}
Note that the above derivation implicitly assumed $N_B \geq n$.
For $N_B < n$,  the operator $\hat{b}^n$ attempts to annihilate more quanta than are present in the initial state. Consequently:
\begin{equation}
    \hat{b}^n \ket{N_B} = 0 \implies A_+ = 0
\end{equation}
Therefore, the system can only undergo a depletion of mode A  (creating B-bosons), and the $A_-$ amplitude remains unchanged. 
After substituting $A_+ = 0$ into our variance formula, we obtain 
\begin{equation}
    \Delta{\hat{n}_B}(t) = n^2 t^2 g_n^2 N_A \frac{(N_B + n)!}{N_B!}
\end{equation}
In this regime, the variance is driven entirely by the one-directional flow of energy into mode B.
Therefore, in the $n\gg 1$ limit at a fixed $N_A$ and $N_B$, since $g_n^2 (N_B + n)!/N_B! \sim n g^2$ and $t_c \sim 1/g\sqrt{n}$, we deduce $\Delta{\hat{n}_B}(t) \propto n^2 (t/t_c)^2$.

\section{Two-mode Squeezing} \label{app:squeezing}
As noted in the main section, charging the non-linear bosonic quantum battery can generate two-mode squeezing, in addition to QFI. 
One way to motivate the generation of  squeezing is to freeze the dynamics of the A-bosons, such that $\hat a \to \expval{\hat a}$ and the non-linear term ($\hat H_\mathrm{AB}$) reduces to an $n$-th order self-interaction for the B bosons.
It is known that in the $n =2$ limit one-mode squeezed states are generated during time evolution under such Hamiltonians~\cite{squeezeoperator}.
Here, we are presented with a stronger non-linearity at a general $n$ which also generates squeezing if the initial state is a coherent state and not a Fock state~\cite{hillery1984squeezing}.
Thus, both the existence and the degree of squeezing  is sensitive to the chice of the initial state.
We note that recent experiments on trapped ions have implemented Hamiltonians with $n>2$ non-linearities to realize  higher-order squeezing.~\cite{nonlinearsqueezing}, which implies that the discussion to follow is  experimentally relevant.

Motivated by prior works, here we consider an initial state $\ket{\psi_i}$ to be a coherent state where
$\ket{\psi_i} =  \ket{\alpha}_A \ket{\beta}_B$ with $a \ket{\alpha} = \alpha \ket{\alpha}$ and $b\ket{\beta} = \beta \ket{\beta}$ and demonstrate the generation of non-trivial two-mode quantum squeezing during the charging cycle of the non-linear bosonic battery.
Due to the connection established in section~\ref{app:collective-spin}, the results can also be interpreted through the lens of a generalized $n$-twist Hamiltonian.
Below, we first present a summary of our findings and then discuss the details in the subsequent subsections.

\subsection{Summary of results}\label{app:summarysqueezing}
To readily see the generation of squeezing, following Ref.~\cite{hillery1984squeezing}, we perform a  short time-evolution of $\ket{\psi_i}$ under $\hat H = \hat H_\mathrm{A} + \hat H_\mathrm{B} + \hat H_\mathrm{AB}$, which leads to  time-dependent variances of $B$-mode quadratures (see Sec.~\ref{app:short-t} for details),
\begin{align}
& \mbox{var}\{\hat x_b(t)\} = \frac{1}{4} - \frac{n(n-1)}{2} g_n t \Im{\alpha^* \beta^{n-1}} + \order{t^2}, \\
& \mbox{var}\{\hat p_b(t)\} = \frac{1}{4} + \frac{n(n-1)}{2} g_n t \Im{\alpha^* \beta^{n-1}} + \order{t^2}.
\end{align}
Here, $\hat x_b = (\hat b + \hat b^\dagger)/2$ and $\hat p_b = (\hat b - \hat b^\dagger)/(2i)$ are conjugate $B$-mode quadratures that evolve in time due to the non-linear terms in $\hat H_\mathrm{AB}$. 
In contrast, variances of analogously defined $A$-mode quadratures, $\hat x_a(t)$ and $\hat p_a(t)$, remain unaltered up to $\order{t^2}$ owing to the linear dependence of $ \hat H_\mathrm{AB}$ on $\hat a$. 
Since a variance below $1/4$ of a bosonic quadrature indicates squeezing~\cite{squeezereview}, assuming $\Im{\alpha^* \beta^{n-1}} > 0$, we deduce that the quadrature $\hat x_b$ gets squeezed in the short-time limit with the squeezing improving as $n$ increases. 
We emphasize that this result is only valid in the short time limit where $|g_n t| \ll 1$.

\begin{figure}[!t]
\centering
\includegraphics[width=0.5\columnwidth]{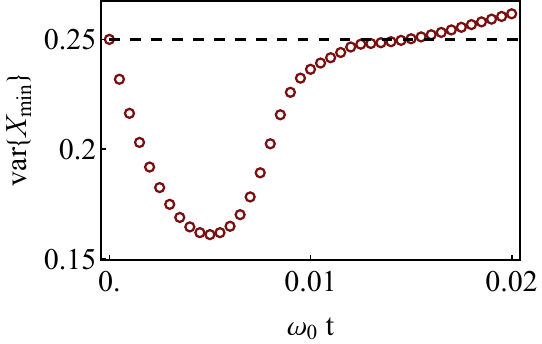}
\caption{Time-evolution of the variance of the numerically optimized squeezed two-mode operator $X_\textrm{min}$. 
Here, we have fixed $(n, \omega_0, g_n) = (4, 1, 1/\sqrt{6})$ and chosen an initial coherent state with $(N_a, N_b) = (|\alpha|^2, |\beta|^2) = (16, 4)$.
}
\label{fig:X}
\end{figure}

We use exact-diagonalization numerics to go beyond the short-time limit as the exact dynamical solution for these highly non-linear Hamiltonians is notoriously challenging. Through our numerics, we verify that squeezing indeed occurs and survives for finite time periods when $n>2$. In order to numerically identify the maximally squeezed quadrature at an arbitrary $t$, we define a generalized two-mode quadrature~\cite{meystre2007elements, sur2025amplified} 
\begin{align}
\hat X_{\theta, \phi, \eta} = \frac{1}{2} \qty[ e^{i\phi} \qty(\cos\theta ~ \hat a + e^{i \eta} \sin\theta ~ \hat b)  + \mbox{h.c.}]
\end{align}
that is continuously parameterized by the three angles, $(\theta, \phi, \eta)$. We numerically optimize over these angles to minimize $\mathrm{var}\{ \hat X_{\theta, \phi, \eta}(t)\}$ at each time $t$.
In Fig.~\ref{fig:X} we show the variance of the maximally squeezed quadrature, $X_\textrm{min}(t)$, for $n=4$ with the initial coherent state parameterized by $(\alpha, \beta) = (-4 i, 2)$.  

We find that the variance of $\hat X_{\text{min}}(t)$ drops below $1/4$ immediately, verifying the squeezing obtained in the short-time limit analytical calculation above. 
The squeezing survives for a substantial time period and hits a minimum at a finite time. 
At this optimal time, we find that $\phi \sim \pi/2$, $\eta \sim 3\pi/2$ and $\theta \lesssim 
\pi/2$. 
We show the time evolution of these parameters and the optimal quadrature for other values of $n$ in Sec.~\ref{app:sqoptimal}. 
The value of $\theta$ indicates that most of the squeezing comes from the $B$-mode which we expect since the non-linearity in the charging Hamiltonian is entirely in the $B$-mode. 
However, there is a finite contribution from the $A$-mode to the optimally squeezed quadrature in general, indicating a  
two-mode squeezing. 
We note that, here, the choice of parameters is dictated by numerical tractability and serves as a proof-of-concept to demonstrate the existence of non-trivial squeezing. 
A detailed investigation of the optimal parameters for maximizing squeezing is beyond the scope of this work.

\subsection{Short-time evolution of squeezing} \label{app:short-t} 
Variance of an operator $\hat X$ whose unitary time-evolution is generated by a Hamiltonian $\hat H$ is given by
\begin{align}
\mbox{var}\{\hat X(t)\} = \mbox{var}\{\hat X(0)\} + i t \delta \mbox{var}\{\hat X\} + \order{(\Omega t)^2},
\end{align}
where $\Omega$ is a characteristic energy scale such that $t$ is measured in units of $\Omega^{-1}$, and 
\begin{align}
\delta \mbox{var}\{\hat X\}  = \expval{\qty{[\hat H, \hat X], \delta \hat X }}
\end{align}
with $\delta \hat X =  \hat X - \expval{\hat X}$ and $\{\hat A, \hat B\} =  \hat A \hat B + \hat B \hat A$.
We note that $\expval{\delta \hat X} = 0$.

For $\hat H = \omega_0 \qty[ n  \hat a^\dagger \hat a + \hat b^\dagger \hat b ] 
+ g_n ( \hat a ^\dagger \hat b^n + \mbox{h.c.})$,
we obtain 
\begin{align}
& \qty[\hat H, \hat x_b] = - i \omega_0 \hat p_b + \frac{n}{2} g_n \qty(\hat a^\dagger \hat b^{n-1} - \mbox{h.c.}); 
\qquad  
\qty[\hat H, \hat x_a] = - i n \omega_0 \hat p_a - \frac{1}{2} g_n \qty(\hat b^{n} - \mbox{h.c.}); \nn \\
& \qty[\hat H, \hat p_b] = i \omega_0 \hat x_b + \frac{i n}{2} g_n \qty(\hat a^\dagger \hat b^{n-1} + \mbox{h.c.}); 
\qquad  
\qty[\hat H, \hat p_a] =  i n \omega_0 \hat x_a + \frac{i}{2} g_n \qty(\hat b^{n} + \mbox{h.c.});
\end{align}
In a coherent state of the form $\ket{\psi_0} = \ket{\alpha} \otimes \ket{\beta}$,  $\expval{\delta a^2} = 0$. 
Consequently, $\{ \delta \hat p_a, \delta \hat x_a \} = 0$ and 
\begin{align}
\delta \mbox{var}\{\hat x_a \} = -\expval{ \qty{ i n \omega_0 \hat \delta p_a + \frac{g}{2} \qty(\hat b^n - \mbox{h.c.} ) + i n \omega_0 \expval{\hat p_a}, \delta \hat x_a  } }
= 0.
\end{align}
Similarly, $\delta \mbox{var}\{\hat p_a \} = 0$.
By contrast, 
\begin{align}
& \delta \mbox{var}\{\hat x_b \} = \frac{n}{2} g_n \expval{ \qty{ \hat a^\dagger \hat b^{n-1} - \mbox{h.c.}, \delta \hat x_b }  } = i \frac{n(n-1)}{2} g_n \Im{\alpha^* \beta^{n-1}};  \\
& \delta \mbox{var}\{\hat p_b \} = - i \frac{n(n-1)}{2} g_n \Im{\alpha^* \beta^{n-1}},
\end{align}
where we have used the result,
\begin{align}
\expval{ \qty{\hat a^\dagger \hat b^{n-1}, \delta \hat x_b } } &= \frac{n(n-1)}{2} \alpha^* \beta^{n-1} \\
&= \expval{ \qty{ (\hat a^\dagger \hat b^{n-1})^\dagger, \delta \hat x_b } }^*.
\end{align}

Since $\mbox{var}\{\hat X(0) \} = 1/4$ for $\hat X \in \{\hat x_a, \hat p_a, \hat x_b, \hat p_b \}$, we obtain the results noted earlier in Sec.~\ref{app:summarysqueezing}.

\subsection{Time-evolution of optimal parameters for squeezing}\label{app:sqoptimal}

\begin{figure}[!t]
\centering
\subfloat[]{%
\includegraphics[width=0.33\columnwidth]{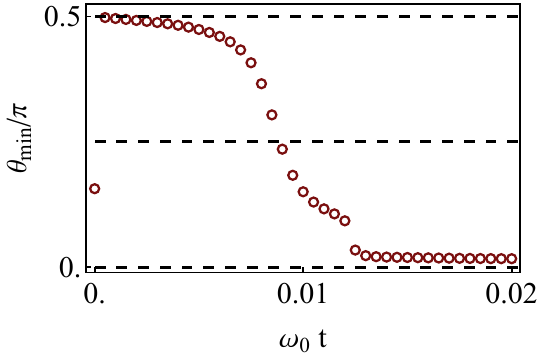}%
}
\hfill
\subfloat[]{%
\includegraphics[width=0.33\columnwidth]{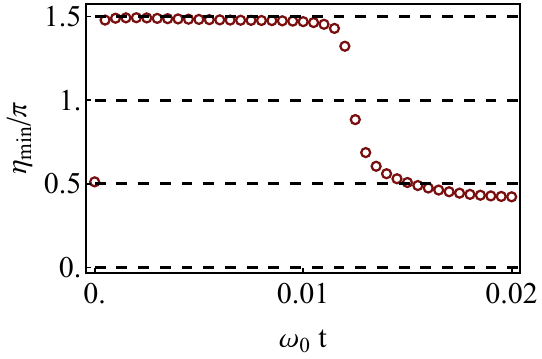}%
}
\hfill
\subfloat[]{%
\includegraphics[width=0.33\columnwidth]{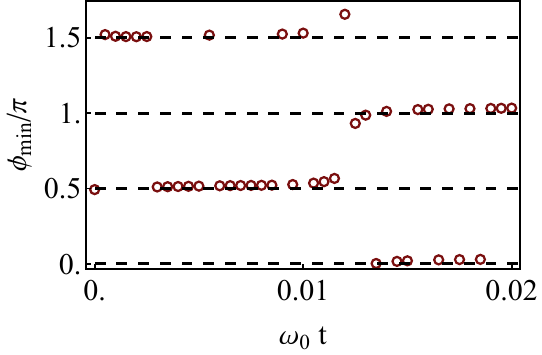}%
}
\caption{Time evolution of the optimal angles for which the variance of the two-mode quadrature $\hat X_{\theta,  \eta, \phi}$ is minimized at $n=4$.} 
\label{fig:angles}
\end{figure}

Here, we include additional details on the determination of the  two-mode quadrature with the lowest variance. 
In Fig.~\ref{fig:angles} we show the time-evolution of the optimal angular parameters which determine the operator $\hat X_{\theta_\mathrm{min}, \eta_\mathrm{min}, \phi_\mathrm{min}}(t)$.
In turn, $\hat X_{\theta_\mathrm{min}, \eta_\mathrm{min}, \phi_\mathrm{min}} (t)$ supports the lowest variance at  $t$, whose time-evolution is presented in Fig.~\ref{fig:X}.

\begin{figure}[!t]
\centering
\subfloat[]{%
\includegraphics[width=0.4\columnwidth]{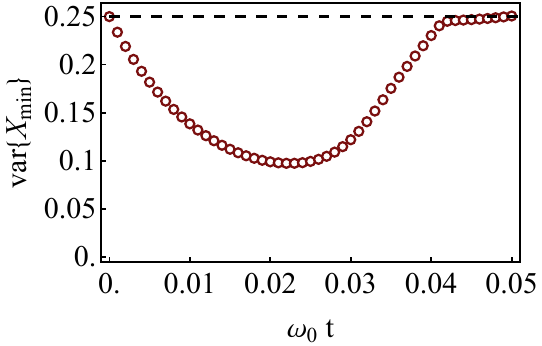}%
}
\hfill
\subfloat[]{%
\includegraphics[width=0.4\columnwidth]{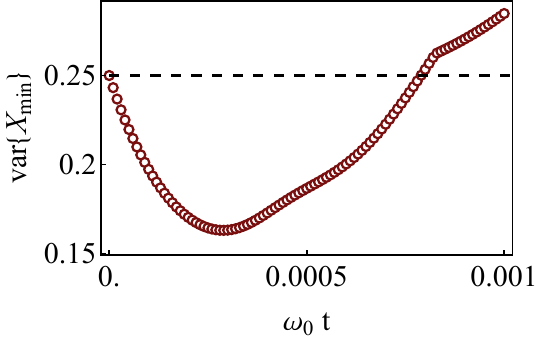}%
}
\caption{Time evolution of the variance of the optimal quadrature $\hat X_{\theta,  \eta, \phi}$ for (a) $n=3$ and (b) $n=6$. The best squeezing is obtained at a finite short time in both cases. 
} 
\label{fig:other-n-1}
\end{figure}

\begin{figure}[!t]
\centering
\subfloat[]{%
\includegraphics[width=0.33\columnwidth]{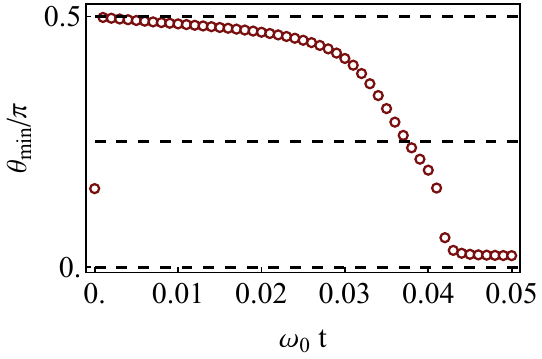}%
}
\hfill
\subfloat[]{%
\includegraphics[width=0.33\columnwidth]{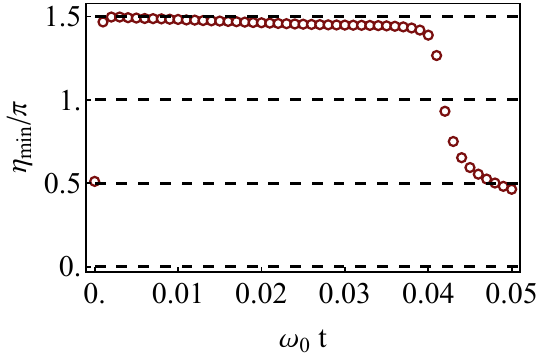}%
}
\hfill
\subfloat[]{%
\includegraphics[width=0.33\columnwidth]{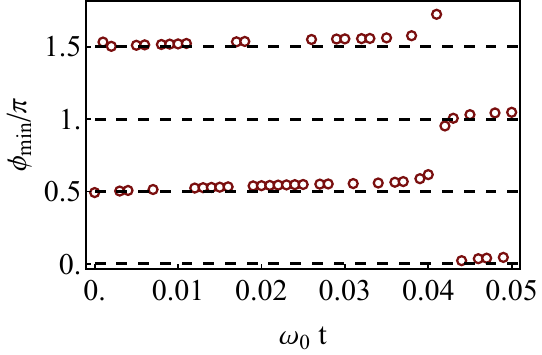}%
}\\
\subfloat[]{%
\includegraphics[width=0.33\columnwidth]{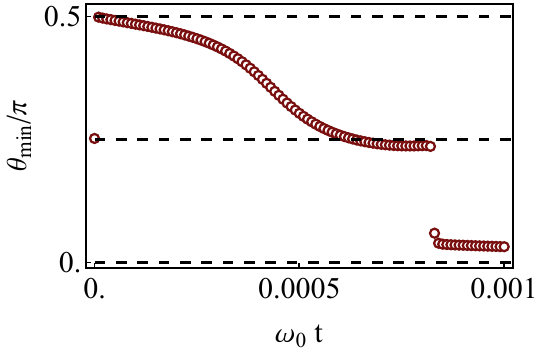}%
}
\hfill
\subfloat[]{%
\includegraphics[width=0.33\columnwidth]{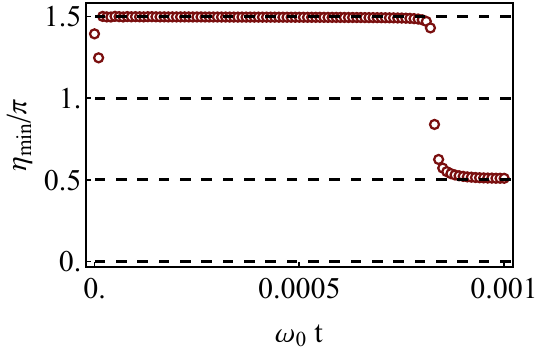}%
}
\hfill
\subfloat[]{%
\includegraphics[width=0.33\columnwidth]{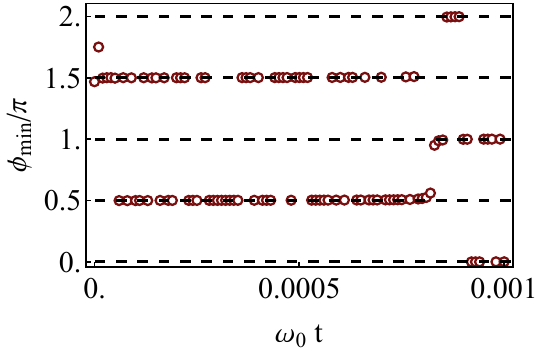}%
}
\caption{Time evolution of the optimal angles for which the variance of the two-mode quadrature $\hat X_{\theta,  \eta, \phi}$ is minimized for (a)--(c) $n=3$ and (d)--(f) $n=6$.} 
\label{fig:other-n-2}
\end{figure}

In Fig.~\ref{fig:other-n-1} we demonstrate the existence of squeezing at  $n = 3$ [subfigure (a)] and $n = 6$ [subfigure (b)]. 
For $n=3$ we used the same initial parameters as the $n=4$ case discussed in the main text, except the choice of $g_n = 1/\sqrt{2}$.
For $n=6$ we have exchanged the values of $|\alpha|$ and $|\beta|$ compared to the cases $n=3, 4$ and chosen  $g_n = 1/\sqrt{5!}$.
In Fig.~\ref{fig:other-n-2} we plot the time evolution of the optimal angles. 

\end{document}